\documentclass{evn2024}
\usepackage{natbib} 
\usepackage{graphicx}
\begin{document}
   \title{New look at old friends}
   \subtitle{EVN imaging of prominent radio-loud active galactic nuclei with extremely large radio--optical positional offsets}

   \author{S.~Frey\inst{1,2,3}
          \and
          O.~Titov\inst{4}
          \and
          A.~Melnikov\inst{5}
          \and
          S.~Lambert\inst{6}
          }

   \institute{Konkoly Observatory, HUN-REN Research Centre for Astronomy and Earth Sciences, Budapest, Hungary
         \and
         CSFK, MTA Centre of Excellence, Budapest, Hungary
         \and
         Institute of Physics and Astronomy, ELTE E\"otv\"os Lor\'and University, Budapest, Hungary
         \and
         Geoscience Australia, Canberra, Australia
         \and
         Institute of Applied Astronomy, Russian Academy of Sciences, St. Petersburg, Russia
         \and
         SYRTE, Observatoire de Paris, Paris, France
             }

   \abstract{
When comparing modern fundamental reference frames in the radio (International Celestial Reference Frame) and optical (\textit{Gaia}), a couple of bright radio reference sources appear to have very large radio--optical offsets, from tens up to hundreds of milliarcseconds (mas). The amount of these positional misalignments exceeds the uncertainty of each individual technique by at least an order of magnitude. In most cases, complex and extended radio structure and its time variability, and thus the difficulty in pinpointing the true location of the central engine, is responsible for the large apparent offsets. Sometimes distant parts of the radio structure are not properly detected due to a lack of shorter interferometer baselines. For our 5-GHz very long baseline interferometry (VLBI) experiment using antennas of the European VLBI Network and the enhanced Multi Element Radio Linked Interferometer Network, we selected 10 bright radio-loud active galactic nuclei with extremely large radio--optical offsets. Sensitive imaging involving a wide range of projected baseline lengths, as well as phase-referencing to nearby sources shed light on the possible causes of positional inconsistencies. Here we show results for 3 selected sources from this project.
   }

   \maketitle
%

\section{Introduction}

The International Celestial Reference Frame (ICRF) is important for astronomy, space navigation, and numerous astrophysical applications \citep{2021FrASS...8....9K}. The International VLBI Service for Geodesy and Astrometry \citep[IVS,][]{2017JGeod..91..711N} is responsible for running regular very long baseline interferometry (VLBI) sessions to produce more observational data around the sky. However, the ICRF, as any quickly-developing scientific product, needs regular verification and update to address all modern challenges. The most recent ICRF realization, known as ICRF3 \citep{2020A&A...644A.159C}, reaches an accuracy of $\sim30\,\mu$as in both coordinate directions (right ascension and declination). In the optical, the \textit{Gaia} \citep{2016A&A...595A...1G,2023A&A...674A...1G} measurements provide the fundamental reference frame. These two frames were intensively compared by different groups \citep[e.g.][]{2016A&A...595A...5M,2017MNRAS.467L..71P,2017A&A...598L...1K,2019ApJ...871..143P,2024A&A...684A.202L}. In spite of the fact that the majority of the common reference sources demonstrate a satisfying positional alignment, it was found that a fraction of the reference sources display large ($\sim 10-500$~mas) positional offsets between the radio and optical coordinates which are far beyond the formal positional uncertainty ($\sim1-5$~mas). Such offsets require comprehensive explanation. In addition, these large offsets are not necessarily constant over a time range of $10-20$ years. Although the majority of radio reference frame sources demonstrate a moderate astrometric instability, about $1-2$~mas \citep[e.g.][]{2011A&A...529A..91T}, we have recently discovered several radio sources \citep{2020RNAAS...4..108T,2021RNAAS...5...60F,2022MNRAS.512..874T} whose apparent position has changed by $20-130$~mas or even more on a time scale of $5-20$ years \citep[see also a recent study of][]{2024ApJS..274...28C}. Such unforeseen dramatic positional changes cause serious problems for geodetic VLBI data analysts because none of the standard software packages is designed to handle them. The data for a whole $24$-h experiment may be totally lost if such objects are not handled properly during the data processing. 

From astrophysical point of view, there is no universal mechanism responsible for this type of astrometric instability. Some radio sources have a typical compact core--jet structure and the apparent motion could be caused by sudden brightening of jet components moving with apparent superluminal speed. Another possible mechanism is the variability of the core flux density within a complex extended radio source. An alternative scenario is a compact symmetric object (CSO) that presents two symmetric lobes (with hotspots) separated on scales of up to $\sim100$~mas beside a weak or often even invisible radio core \citep[e.g.][]{2012ApJ...760...77A,2024ApJ...961..240K}. In this case, the astrometric instability is likely to be caused by the variations of flux density in the lobes.

In this conference contribution, we highlight some of the preliminary results of our VLBI imaging observations performed with the European VLBI Network (EVN), also involving the antennas of the enhanced Multi Element Radio Linked Interferometer Network (e-MERLIN), at $5$~GHz frequency in 2022 March (project code: ET048, PI: O. Titov). We targeted a total of $10$ prominent radio-loud active galactic nuclei (AGN) that show a large radio--optical positional offset and a hint of significant astrometric instability. The sources were selected from the literature \citep{2017ApJ...835L..30M,2017MNRAS.467L..71P} and from the Paris Observatory database. Historical VLBI images are sparsely available for these sources in the Astrogeo database (astrogeo.org), mostly at S and X bands, i.e. around $2$ and $8$~GHz \citep{2024arXiv241011794P}, and in the literature, but those are dominantly derived from short snapshot VLBI observations with limited sensitivity. However, the archival data are essential for comparison with the new EVN images, to determine whether there
is any sign of significant structural variation behind the extreme astrometric instability.


\section{Observations and data reduction}

The $5$-GHz EVN and e-MERLIN observations were done in 3 segments (experiment ET048A with 7 target sources starting on 2022 Mar 9, ET048B with 2 target sources on 2022 Mar 10, and ET048C with a single target on 2022 Mar 14). The data were recorded in four intermediate frequency channnels, each containing sixty-four $0.5$-MHz-wide spectral channels, in left and right circular polarizations. The data were correlated at the Joint Institute for VLBI ERIC (Dwingeloo, The Netherlands) with $2$-s integration time. 
About $1.5$-h long observing blocks were scheduled for each target source and its phase-reference calibrator. The participating radio telescopes that provided data were Jodrell Bank Mk2 (United Kingdom), Effelsberg (Germany), Medicina (Italy), Noto (Italy; only in ET048C), Onsala 25-m (Sweden), Tianma (China), Nanshan (China), Toru\'n (Poland), Yebes (Spain), Irbene 16-m (Latvia; except for ET048C), Hartebeeshoek (South Africa), Cambridge, Darnhall, Defford, Knockin, and Pickmere (e-MERLIN stations in the United Kingdom).   

The correlated visibility data were calibrated using the NRAO Astronomical Image Processing System \citep[\textsc{AIPS},][]{2003ASSL..285..109G} in a standard way \citep[e.g.][]{1995ASPC...82..227D}. The calibrated data were then exported to the \textsc{Difmap} software package \citep{1994BAAS...26..987S} for hybrid mapping. Both direct fringe fitting \citep{1983AJ.....88..688S} and phase referencing to nearby (within $\sim 2^{\circ}$) bright ICRF calibrator sources were performed for the targets in \textsc{AIPS}. The former procedure provides more sensitive images, at the expense of losing absolute astrometric information, as it associates the brightness peak with the phase centre. On the other hand, phase referencing allows us to preserve the true sky position of the brightness peak, determined with respect to the nearby calibrator. 


\section{Results and discussion}

We present images of 3 target sources as examples, and discuss the possible causes of their radio--optical positional differences. 

\begin{figure}
\centering
\includegraphics[width=\columnwidth]{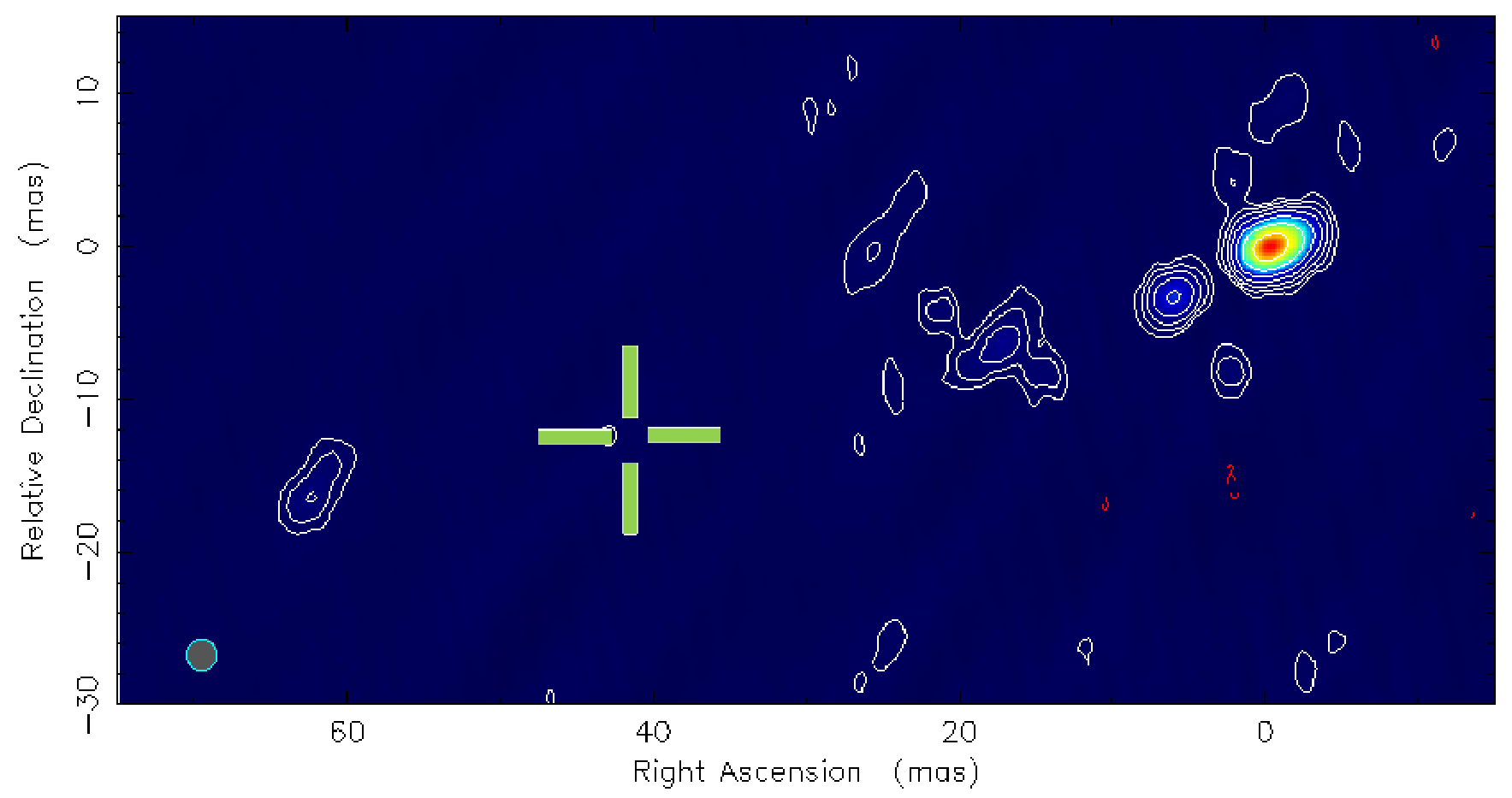}
      \caption{Naturally-weighted 5-GHz EVN image of PKS~1200+045 from 2022 Mar 9 (ET048A). The crosshairs indicate the \textit{Gaia} DR3 position which is more than $40$~mas away from the radio peak whose position is determined with phase referencing to J1202+0235 (separated by $1.65^{\circ}$). The peak brightness is $156$~mJy\,beam$^{-1}$, the lowest contours are drawn at $\pm 1.7$~mJy\,beam$^{-1}$ ($\sim 3.7\sigma$ image noise), the positive contour levels increase by a factor of 2. The half-power width of the circular Gaussian restoring beam is $2$~mas, as indicated in the lower-left corner.}
         \label{fig:J1203}
\end{figure}

\textbf{PKS~1200+045} (J1203+0414) is a bright peaked-spectrum radio quasar at redshift $z=1.224$. Its radio structure seen in our $5$-GHz EVN image (Fig.~\ref{fig:J1203}) is extended to $\sim 60$~mas (corresponding to $\sim 0.5$~kpc projected linear size). It is consistent with earlier VLBI images from the literature. That time it was not clear whether its compact radio structure can be classified as a core--jet \citep{2007A&A...470...97L,2010A&A...518A..23C} or a CSO \citep{2007A&A...470...97L}. The \textit{Gaia} optical position marked by a cross in the image differs significantly from the radio peak position \citep{2017ApJ...835L..30M,2017MNRAS.467L..71P}, lending strong support for the CSO classification. It is one of the typical causes of radio--optical offsets because \textit{Gaia} basically pinpoints the accretion disk emission from around the central supermassive black hole, often invisible in the radio, while the radio emission peak is associated with a bright hotspot in a lobe \citep[see another example in][]{2020MNRAS.496.1811K}.

\begin{figure}
\centering
\includegraphics[width=\columnwidth]{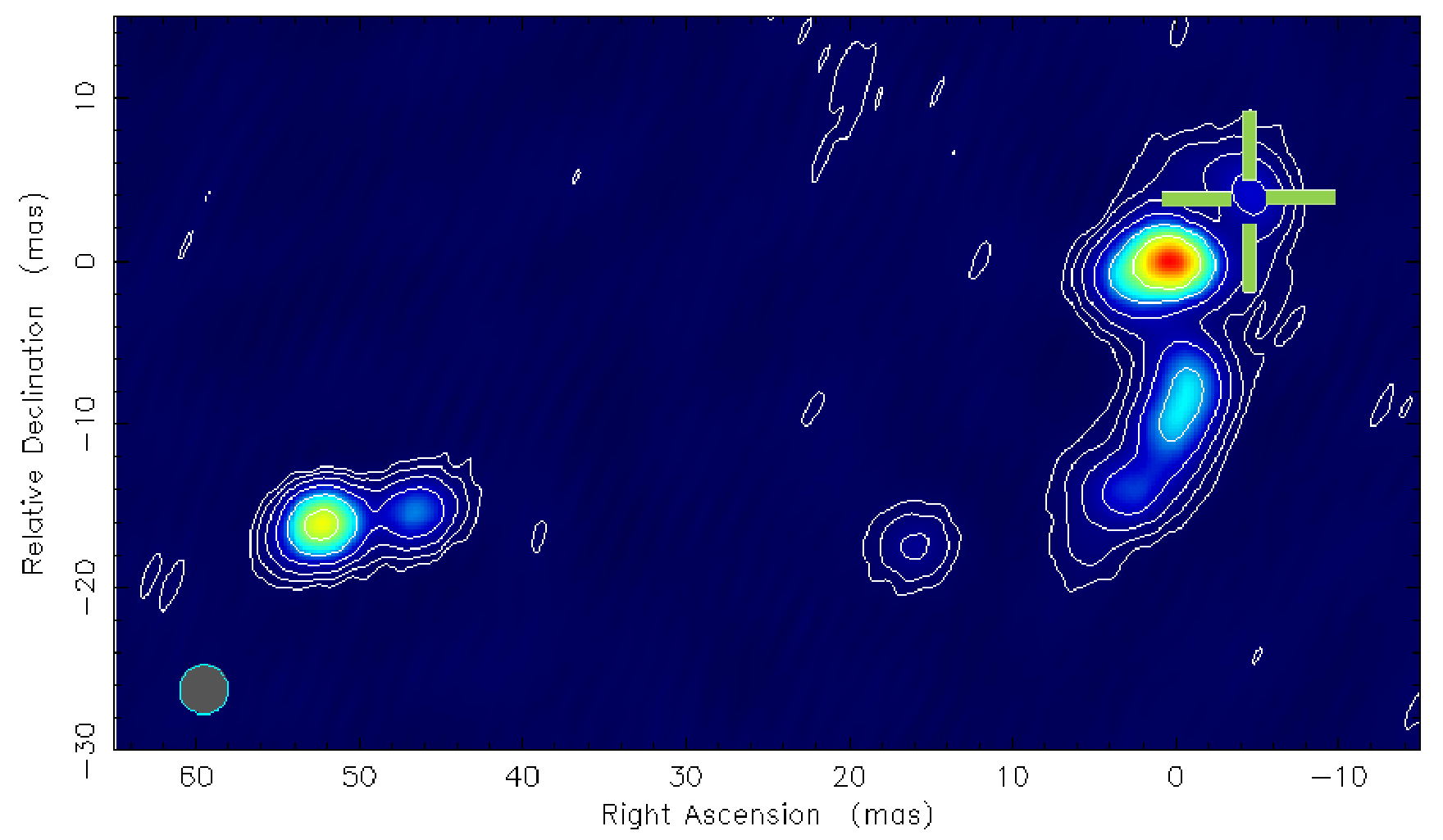}
      \caption{Naturally-weighted 5-GHz EVN image of TXS~1450+641 from 2022 Mar 10 (ET048A). The crosshairs indicate the \textit{Gaia} DR3 position. The radio peak position is determined with phase referencing to J1441+6318 (separated by $1.28^{\circ}$). The peak brightness is $24.9$~mJy\,beam$^{-1}$, the lowest contours are drawn at $\pm 0.4$~mJy\,beam$^{-1}$ ($\sim 3.5\sigma$ image noise), the positive contour levels increase by a factor of 2. The half-power width of the circular Gaussian restoring beam is $3$~mas, as indicated in the lower-left corner.}
         \label{fig:J1451}
\end{figure}

\textbf{TXS~1450+641} (J1451+6357) is a quasar with unknown redshift. Because of its complex radio structure dominated by two bright features \citep{2007ApJ...658..203H}, it was once considered but then refuted as a CSO candidate \citep{2016MNRAS.459..820T}. Indeed, the \textit{Gaia} optical position is not in between two main features but close to the western one (Fig.~\ref{fig:J1451}), although there is a clear offset from the radio brightness peak. The interpretation of this apparently peculiar structure which may involve a turning or bending jet is beyond the scope of this astrometric study. Note that an earlier VLBI image at the same frequency ($5$~GHz) shows the easternmost component as the brightest one \citep{2007ApJ...658..203H}. This reminds us the fact that VLBI astrometric positions can sometimes be influenced by dramatic changes in the brightness distribution of the sources \citep{2022MNRAS.512..874T}.

\begin{figure}
\centering
\includegraphics[width=0.7\columnwidth]{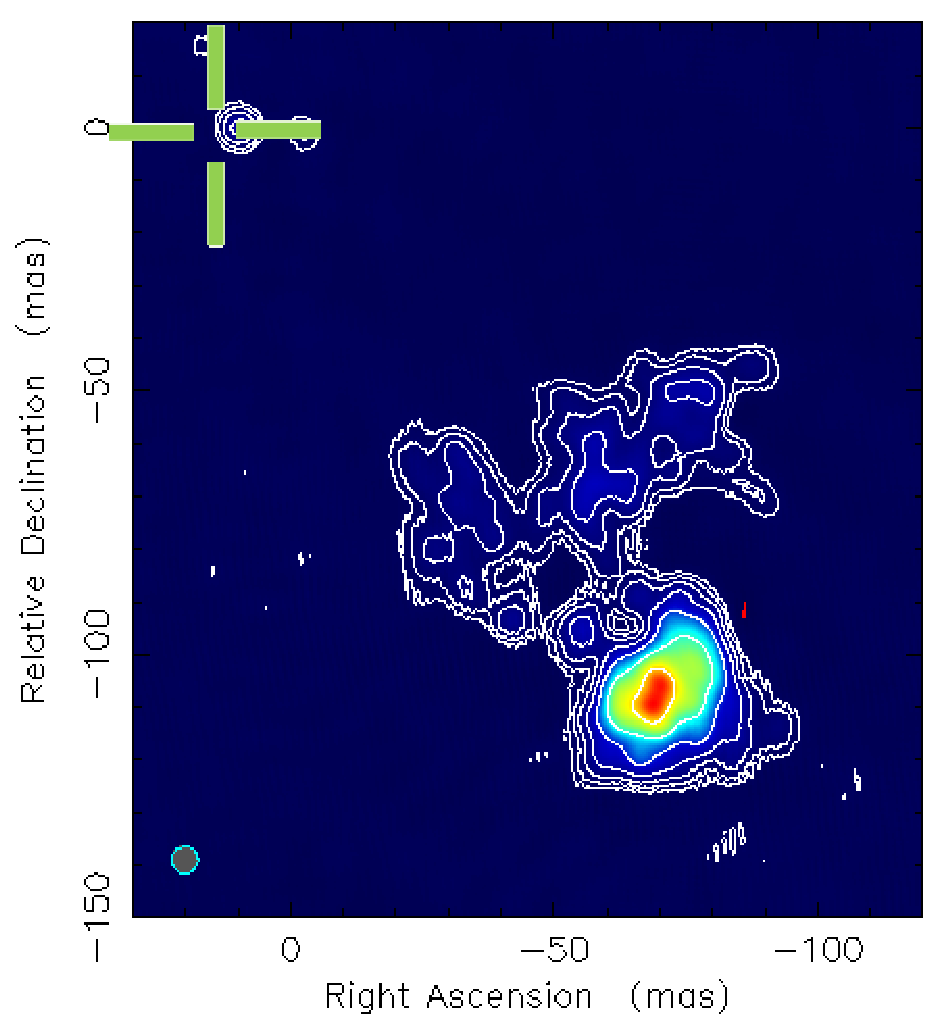}
      \caption{Naturally-weighted 5-GHz EVN image of 3C\,287 from 2022 Mar 14 (ET048C). The crosshairs indicate the \textit{Gaia} DR3 position. The image is phase-referenced to J1333+2725 (separated by $2.34^{\circ}$). The peak brightness is $175$~mJy\,beam$^{-1}$, the lowest contours are drawn at $\pm 2$~mJy\,beam$^{-1}$ ($\sim 4\sigma$ image noise), the positive contour levels increase by a factor of 2. The half-power width of the circular Gaussian restoring beam is $5$~mas, as indicated in the lower-left corner.}
         \label{fig:J1330}
\end{figure}

\textbf{PKS~1328+254} (J1330+2509, also known as 3C\,287) is a bright steep-spectrum quasar ($z=1.055$) extensively studied from the early years of VLBI \citep{1985A&A...143..292F,1989A&A...217...44F,1990A&A...231..333F}, until before 2000 \citep{1998A&A...338..840P}. The extended and complex radio structure seen in the southwestern part of our image (Fig.~\ref{fig:J1330}) was originally interpreted as a helical jet shaped by precession. This, and the identification of the radio core inside were later questioned by several authors \citep{1998A&A...338..840P,1998AJ....115.1295K}. Intriguingly, the VLBI astrometric position of 3C\,287 was displaced by $\sim130$~mas within a very short time, between 2014 and 2017 \citep{2022MNRAS.512..874T}. It was interpreted as caused by a sudden brightening of a new component in the northeastern direction \citep{2022MNRAS.512..874T}. Unfortunately, modern VLBI imaging data were not available in the archives for verifying this scenario. In our new EVN image (Fig.~\ref{fig:J1330}), we do detect this putative component for the first time, even though its present-day flux density ($\sim 20$~mJy) is not particularly large. Most notably, the \textit{Gaia} optical position in the phase-referenced image is close the this component which might have gone through a huge radio outburst in the mid-2010’s. In any case, in the light of new astrometric and imaging VLBI data, the time has arrived for revisiting 3C\,287 for astrophysical interpretation of its radio structure on $\sim10-100$~mas angular scale.

\section{Summary}

Using the EVN and e-MERLIN arrays, we imaged 10 prominent radio-loud AGN occasionally targeted with geodetic/astrometric VLBI observations, and show radio--optical positional mismatch one or two orders of magnitude larger than expected from uncertainties of accurate VLBI and $Gaia$ measurements in the radio and optical, respectively. Our new 5-GHz images can be used for comparison with archival VLBI images obtained mostly at the $2$- and $8$-GHz bands during the past decades, to help explain the large, sometimes variable positional offsets. Phase referencing to nearby ICRF sources provide accurate astrometric registration of the radio images. 

In this contribution, we presented 3 cases as examples. The quasar PKS~1200+045 is reminiscent of a CSO whose optical position may be associated with the accreting central supermassive black hole, while the radio brightness peak coincides with a hotspot in one of the lobes. The quasar TXS~1450+641 has a complex, time-variable core--jet structure. The optical position is located at one end of the elongated radio structure. The new VLBI image of the source PKS~1328+254 (3C\,287) confirms the existence of a compact mas-scale component recently inferred from an apparent ``jump'' of its VLBI astrometric position. This component, never seen before in any other VLBI image, is located $\sim 130$~mas away from the well-known complex, bright radio structure. It is remarkably close to the \textit{Gaia} optical position, suggesting that it is associated with the central engine of the quasar.

The complete analysis of the full sample of 10 sources is being prepared for publication elsewhere.  

\begin{acknowledgements}
The European VLBI Network (www.evlbi.org) is a joint facility of independent European, African, Asian, and North American radio astronomy institutes. Scientific results from data presented in this publication are derived from the following EVN project code: ET048. e-MERLIN is a National Facility operated by the University of Manchester at Jodrell Bank Observatory on behalf of STFC. 
SF acknowledges support from the Hungarian National Research, Development and Innovation Office (NKFIH, grant OTKA K134213) and the NKFIH excellence grant TKP2021-NKTA-64.

\end{acknowledgements}

\bibpunct{(}{)}{;}{a}{}{,}
\bibliographystyle{aa}
\bibliography{et048}

\end{document}